# Mechanochemical bistability of intestinal organoids enables robust morphogenesis

Shi-Lei Xue[1*], Qiutan Yang[2,3,4,5,6*], Prisca Liberali[2,7°], Edouard Hannezo[1°]

[1]Institute of Science and Technology Austria, Am Campus 1, 3400 Klosterneuburg, Austria

[2]Friedrich Miescher Institute for Biomedical Research (FMI), Maulbeerstrasse 66, 4058 Basel, Switzerland

[3]State Key Laboratory of Stem Cell and Reproductive Biology, Institute of Zoology, Chinese Academy of Sciences, 1 Beichen West Road, Chaoyang District, Beijing 100101, China

[4]Institute for Stem Cell and Regeneration, Chinese Academy of Sciences, Jia 3 Hao Datun Rd. Chaoyang District, Beijing 100101, China

[5]Beijing Institute for Stem Cell and Regenerative Medicine, Jia 3 Hao Datun Rd. Chaoyang District, Beijing 100101, China

[6]University of Chinese Academy of Sciences, NO. 19A Yuquan Rd, Shijingshan District, Beijing 100049, China

[7]University of Basel. Petersplatz 1, 4001 Basel, Switzerland

* Equal contributions

° Corresponding authors: prisca.liberali@fmi.ch; edouard.hannezo@ist.ac.at

**Abstract:** How pattern and form are generated in a reproducible manner during embryogenesis remains poorly understood. Intestinal organoid morphogenesis involves a number of mechanochemical regulators, including cell-type specific cytoskeletal forces and osmotically-driven lumen volume changes. However, whether and how these forces are coordinated in time and space via feedbacks to ensure robust morphogenesis remains unclear. Here, we propose a minimal physical model of organoid morphogenesis with local cellular mechano-sensation, where lumen volume changes can impact epithelial shape via both direct mechanical (passive) and indirect mechanosensitive (active) mechanisms. We show how mechano-sensitive feedbacks on cytoskeletal tension generically give rise to morphological bistability, where both bulged (open) and budded (closed) crypt states are possible and dependent on the history of volume changes. Such bistability can explain several paradoxical experimental observations, such as the importance of the timing of lumen shrinkage and robustness of the final morphogenetic state to mechanical perturbations. More quantitatively, we performed mechanical and pharmacological experiments to validate the key modelling assumptions and make quantitative predictions on organoid morphogenesis. This suggests that bistability arising from feedbacks between cellular tensions and fluid pressure could be a general mechanism to allow for the coordination of multicellular shape changes in developing systems.

**Introduction**

Embryos are sculpted by a variety of physical forces (Heisenberg and Bellaïche, 2013; Collinet and Lecuit, 2021; Goodwin and Nelson, 2021), such as active tensions from the cellular cytoskeleton (Lecuit and Lenne, 2007), and compressive stresses from external physical constraints or differential tissue growth (Li et al., 2011; Hannezo et al., 2011; Savin et al., 2011; Shyer et al., 2013, Karzbrun et al., 2018, Trusko et al, 2020). Furthermore, hydrostatic pressure forces are emerging as important regulators for supracellular morphogenesis (Saias et al., 2015, Nelson et al., 2017; Dasgupta et al., 2018; Chan et al., 2019; Dumortier et al., 2019, Yang et al, 2021; Munjal et al., 2021; Chartier et al., 2021, Le Verge-Serandour et al, 2021, Stokkermans et al, 2022, Huljev et al, 2023, Schliffka et al, 2023). Interestingly, even though physical forces have been established to play key functional roles during morphogenesis, it is still not clear whether and how they are coordinated with each other and with concomitant biochemical signalling events such as morphogen gradients and cell fate specification (Hannezo and Heisenberg, 2019; Zinner et al., 2020; Pinheiro et al., 2022; Valet et al., 2022). Indeed, cells and tissues typically experience large noise both at the biochemical and mechanical levels, begging the question of how morphogenesis can occur in a robust and reproducible manner (Menshykau et al., 2014; Haas et al., 2018; Staddon et al., 2019; Niwayama et al., 2019; Erzberger et al., 2020; Tsai et al., 2020; Guo et al., 2022; Fabrèges et al., 2023).

The development of the vertebrate intestine represents a prototypical example of a complex morphogenetic sequence resulting in a highly stereotypical folded configuration, which is necessary to supply adequate surface area for nutrient absorption (Savin et al., 2011; Shyer et al., 2013). The epithelium of the small intestine is organized in a folded monolayer with highly curved invaginations (crypts), where stem cells reside, and large finger-like protrusions (villi) into the intestinal lumen consisting of differentiated cells (Shyer et al., 2013; Sumigray et al., 2018). To overcome the limited accessibility of internal tissues, intestinal organoids have emerged as an ideal *in vitro* self-organized model system amenable to live-imaging and mechano-chemical perturbations (Serra et al., 2019; Tallapragada et al., 2021; Pérez-González et al., 2021; Yang et al., 2021), as they adopt shapes as well as cell fate distributions highly reminiscent of *in vivo* tissue organization (Sato et al., 2009; Serra et al., 2019) (Fig. S1A-C).

Organoid morphogenesis is organized by fate-dependent forces from crypt and villi regions, respectively actomyosin-driven apical constriction and lumen osmotic forces (Hartl et al., 2019; Yang et al., 2021; Pérez-González et al., 2021) (Fig. 1A and Supplementary Video 1).

However, whether these two key mechanical events are independently regulated or intrinsically coupled at the level of an entire organoid remains unclear. Interestingly, although lumen volume reduction at the onset of morphogenesis was shown to be critical for crypt budding (Fig. 1A-B and Supplementary Video 1-2), once crypts have reached the budded stage, their morphologies become unaffected by lumen volume increases (Fig. 1C and Supplementary Video 3) (Yang et al., 2021), suggesting that the relative timing of different mechanical events is crucial for morphogenesis. We reasoned that this feature, not captured by current theoretical models, could act as a source of robustness for morphogenesis, as it would result in the crypt shape becoming insensitive to mechanical fluctuations. Altogether, these observations call for a more systematic understanding of how different types of forces acting at different scales are coordinated in space and time during morphogenesis.

Here, we propose a biophysical theory for the coordination of mechano-osmotic forces driving intestinal organoid morphogenesis. Via an analytically tractable 3D vertex model, we find that the phase diagram for organoid morphologies contains a mechanically bistable region: both open (bulged) and closed (budded) crypt configurations are possible for a specific organoid volume, and the ultimate morphological outcome depends on the history of the system. Biochemical bistability has been proposed as a key source of robustness for stem cell fate determination by allowing cells to both commit to a given fate upon an inducing signal, and retain this fate after the signal is removed (Alon et al, 2007; Zhao et al, 2022). We reasoned that lumen pressure could play the role of an inducing signal in multicellular systems, with mechanical bistability resolving the apparent experimental paradox of lumen inflation causing opposite outcomes at different time points (Fig. 1D). We also find that the regime of parameters allowing for bistability, and thus robust morphogenesis, is drastically enhanced by considering a mechano-sensitive coupling between actomyosin tension and lumen pressure, which we verified experimentally and allowed us to quantitatively recapitulate organoid morphogenesis.

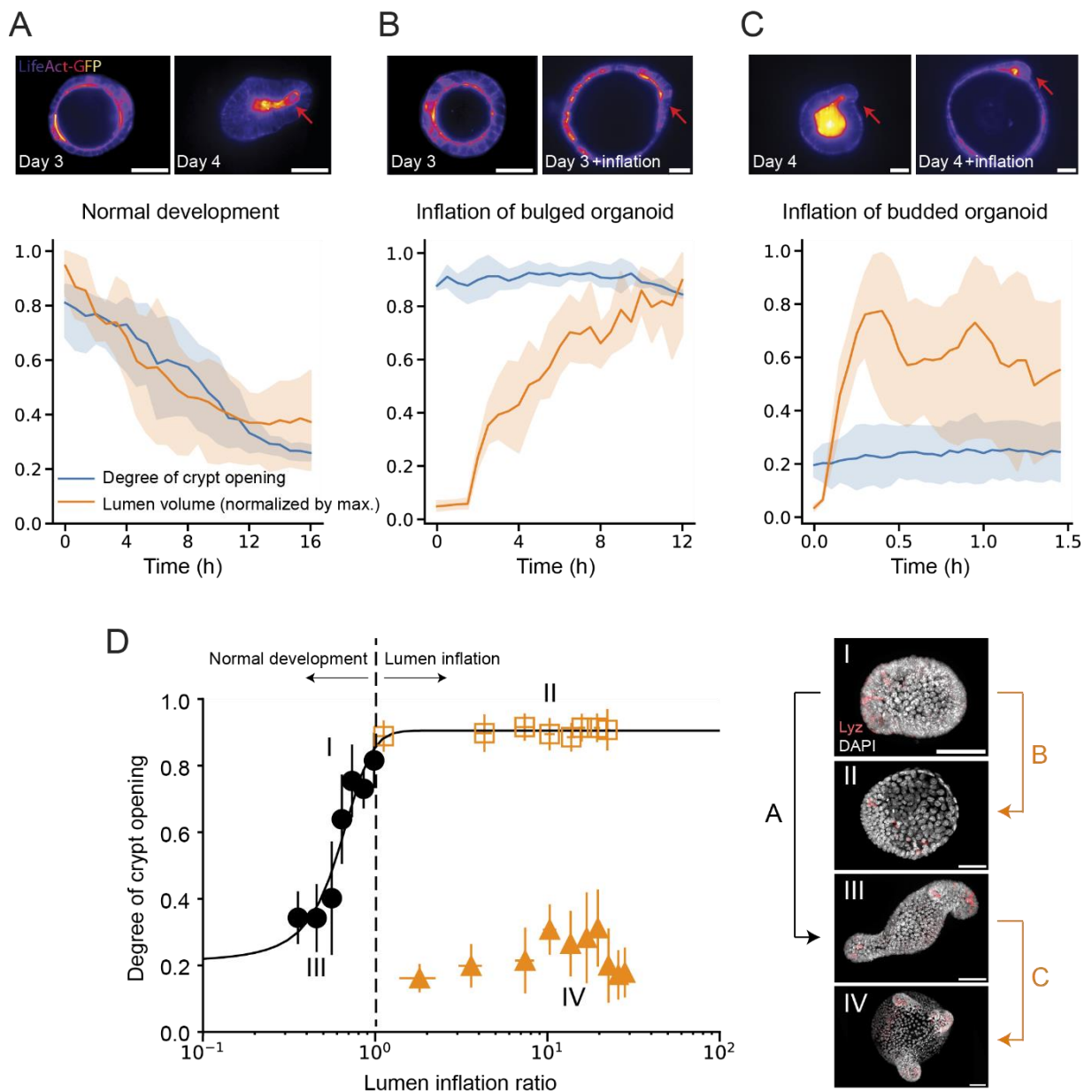

**Figure 1: Crypt morphology depends on lumen volume changes in a history-dependent manner. A-C.** Top, representative time-lapse recordings of the normal development (A) of intestinal organoids, as well as lumen inflation (performed by PGE treatment) of both bulged (B) and budded (C) organoids; bottom, corresponding degree of crypt opening and lumen volume (normalized by maximal value) as a function of time (sample number $N = 4$ in each scenario). **D.** Left, bistable relationship between the degree of crypt opening and lumen inflation ratio: early lumen inflation (B: I→II) impairs long-term crypt budding morphogenesis, while lumen inflation after normal development (C: III→IV) has negligible effect. Right, representative snapshots of 4 types of crypt states: I, Day 3.5 bulged organoid; II, Day 4 organoid that remained bulged due to long-term lumen inflation; III, Day 4 budded organoid (unperturbed); and IV, Day 4 budded organoid after lumen inflation. Images are maximum z-projection of organoids with DAPI and Lyz staining. Scale bars in all images: 50 μm. Error bars represent SD.

**Theory of bistable crypt morphology as a function of lumen volume**

We theoretically describe an intestinal organoid as a closed epithelial monolayer with two different regions, crypt and villus, which encapsulates an incompressible fluid lumen. We have previously shown that such two-region 3D vertex model, which captures the effects of cell-scale active forces on tissue-scale deformation, can accurately predict organoid morphologies (Yang et al., 2021). In the model, we consider a single cell with three surface tensions $\Gamma_a$, $\Gamma_b$ and $\Gamma_l$, arising from cell-cell adhesion and actomyosin-induced tension along the cell membrane (Fletcher et al., 2014; Hannezo et al., 2014; Alt et al., 2017; Rozman et al., 2020), and three surface areas $A_a$, $A_b$, and $A_l$, with subscripts $a$, $b$, and $l$ respectively representing apical, basal, and lateral surfaces/domains (Fig. 2A). Then the free energy of a single cell reads:

$$f = \Gamma_a A_a + \Gamma_b A_b + \frac{1}{2}\Gamma_l A_l.$$

When considering that crypt and villus regions can each have different mechanical properties, the possible morphologies of an organoid depend on four dimensionless parameters (see SI Note for details): i) the relative size of the crypt $\varphi$ (i.e. the ratio of crypt cell number to total number), ii) normalized organoid volume $v$, iii) in-plane tension ratio $\alpha = \frac{(\Gamma_a+\Gamma_b)_c}{(\Gamma_a+\Gamma_b)_v}$ (with subscripts c and v denoting crypt and villus tissues) and iv) differential tension between the apical and basal sides of crypt cells $\sigma_c = \frac{1}{2}\left(\frac{\Gamma_a-\Gamma_b}{\Gamma_l}\right)_c \sqrt{\frac{4\pi}{N_t}}$, which causes the crypt to have a preferred curvature.

Previously, we considered apical Myosin in crypts as the key parameter (Yang et al., 2021), which changes both in-plane tension $\alpha$ and differential tension $\sigma_c$. However, more complex regulatory patterns of Myosin can be observed experimentally, including basal crypt actomyosin relocation during lumen inflation (Fig. S2 and Supplementary Video 4). When investigating theoretically the separate consequences of differential tension $\sigma_c$ and in-plane tension $\alpha$, we found that in specific regions of the phase diagram (Fig. 2A), crypt morphology can show two stable configurations (either open or closed shape) for the same lumen volume, and thus displays morphological bistability. To gain insights into this theoretical phenomenon, we consider the limit of large lumen volume, in which case the dependency of the total mechanical energy of the organoid on crypt shape (represented by crypt opening angle $\theta_c$) can

be derived analytically:

$$\Delta F(x) = \left[1 - \sigma_c \left(\frac{1+x}{2\varphi}\right)^{\frac{1}{2}}\right]^{\frac{2}{3}} \left(1 - \frac{1-x}{2\alpha}\right)^{\frac{1}{3}},$$

where $x = \cos(\theta_c)$. Depending on differential tension $\sigma_c$, three scenarios exist (Fig. 2B): (i) for small $\sigma_c$, the crypt epithelium will stay open/bulged; (ii) for intermediate differential tension $\sigma_c$, two local energy minima exist, indicating that both open and closed configurations are possible; (iii) for large enough $\sigma_c$, the crypt will always be budded. Next, we look at the morphological evolution of organoid shape upon changes of lumen volume. The "degree of crypt opening", described as $\theta_c/(\pi - \theta_v)$, is employed to quantify crypt morphology (Fig. 1 and 2C), ranging from 0 to 1, with 0 standing for the budded shape (with crypt and villus fully closed) and 1 for a spherical organoid. For small differential tension (e.g. $\sigma_c = 0.1$), crypt morphology evolves continuously and monotonously with lumen volume changes: the crypt opens (or closes) on lumen inflation (or shrinkage) (characterized by path $A \to D$). With intermediate differential tension (e.g. $\sigma_c = 0.15$), the crypt will gradually close up with lumen shrinkage until fully closed, however, the morphological evolution is no longer reversible as this closed crypt cannot be opened by further volume inflation (path $B \to C$).

Such morphological hysteresis – or path-dependent evolution – is a classical feature of bistable systems (Alon et al, 2007), and is qualitatively consistent with our data (Fig. 1). Starting from a bulged/open crypt, increasing differential tension at constant volume drives an entry into the bistable region from the "bulged" configuration and crypts are thus prone to stay bulged (path $A \to C$) – mirroring our data that impairing volume decrease during morphogenesis (day 3 to 4) impairs crypt budding (Fig. 1B and Supplementary Video 2). On the other hand, if differential tension increase occurs at the same time as decreasing lumen volume (path $A \to B$), organoids can reach the monostable budded state, consistent with our data on normal organoid morphogenesis (Fig. 1A and Supplementary Video 1). Crucially, subsequently increasing lumen volume back to its initial value once crypts are already budded (path $B \to C$) results in organoids entering the bistable region from the "budded" configuration, and thus remaining in this state, as in our inflation data (Fig. 1C and Supplementary Video 3), despite reaching the same final volume as in the first path $A \to C$.

This suggests that the experimental system could exhibit morphological hysteresis and bistability as in the model, although this needed to be systematically tested. Before turning to more quantitative comparison between data and theory, we proceeded to understand better the mechanical origin of bistability in the model.

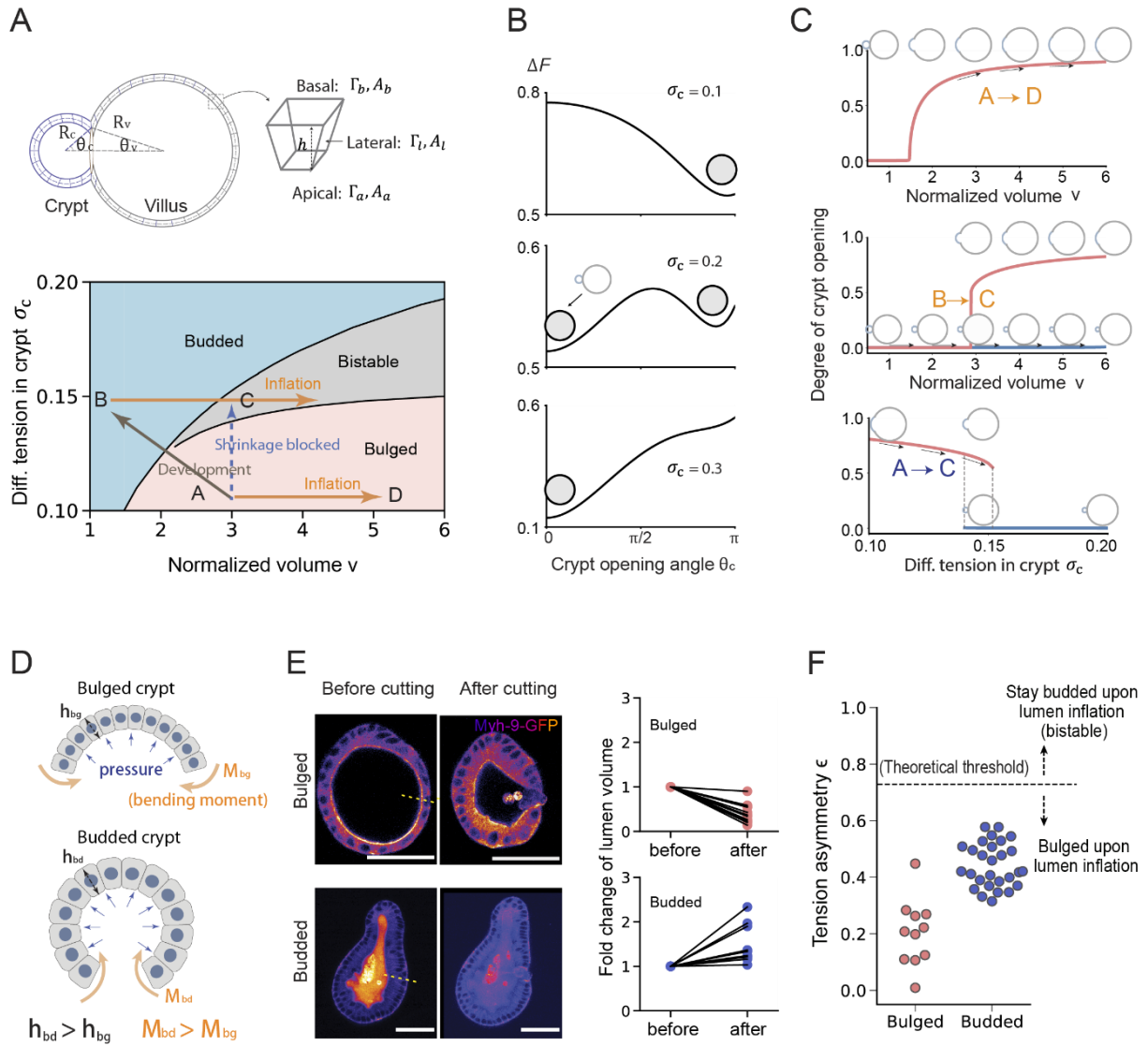

**Figure 2: Theory of bistable crypt morphology switched by lumen volume. A**. Schematic of 3D vertex model (top) and phase diagram of crypt morphology (bottom) as a function of crypt differential tension $\sigma_c$ and normalized volume $v$. In the schematic, $R_c$ and $R_v$ are respectively radii of curvature in crypt and villus; $h$ is cell height/tissue thickness (see Supplementary Theory Note for details). **B**. Three possible energy landscapes that control crypt morphology. Local energy minima are crypt equilibrium states, with $\theta_c = 0$ the budded (or closed) crypt shape and $\theta_c > 0$ the bulged (or open) shape. **C**. Evolution of crypt shape with varying lumen volume at constant differential tension $\sigma_c = 0.1$(top) and $\sigma_c = 0.15$ (middle), and evolution with varying crypt differential tension at constant volume $v = 3$ (bottom). Arrows are different paths shown in phase diagram A. **D**. Schematic of the mechanism of curvature-thickness feedback in crypt epithelium: epithelial bending increases its thickness $h$ and corresponding active bending moment $M \sim \sigma_c h$, which in turn facilitates crypt budding. **E.** Fold changes of lumen volume before and after villus tissue breakage, in both bulged ($N = 11$) and budded ($N = 9$) organoids. **F.** Tension asymmetry of crypt apical vs. basal surfaces in bulged ($N = 11$) and budded ($N = 28$) organoids, and theoretical threshold (dashed line) for bistability due to curvature-thickness feedback. Scale bars in all image: 50 μm.

**Mechanical origin of morphological bistability**

Our 3D vertex model of intestinal organoids is conceptually related to the classical Helfrich elastic theory for lipid vesicles (Jülicher and Lipowsky, 1996, Zimmerberg and Kozlov, 2005), except for the key difference of morphological bistability observed above. However, we reasoned that unlike lipid membranes, epithelia have a comparatively large thickness which changes with active tensions and deformations (Yang et al., 2021; Pérez-González et al., 2021; Messal et al., 2019, Recho et al, 2020, Iber et al, 2022, Khoromskaia et al, 2023).

During crypt morphogenesis, the out-of-plane deformation driven by differences in apico-basal tension tends to increase the epithelial thickness as crypt curvature increases, (Krajnc and Ziherl, 2015; Rozman et al., 2020; Yang et al., 2021; Luciano et al., 2021), and the driving force ("active bending moment") $M$ arising from differential tension is also proportional to epithelial thickness (i.e. $M \sim (\Gamma_a - \Gamma_b)h$, Fig. 2D). Therefore, the epithelial thickness and the out-of-plane deformation can have positive feedback: the bending tends to thicken the epithelium, which in turn enhances $M$ and thus facilitates out-of-plane deformation. This effect saturates, since thicker tissues also have higher bending stiffness. To test this hypothesis computationally, we examined an alternative model where crypt thickness is kept constant, and confirmed that bistability was absent (Fig. S3F). Overall, purely mechanical feedbacks can thus provide a potential explanation for crypt bistability, and thus the history-dependent effects of lumen volume changes that we observed experimentally.

Next, we aimed to quantitatively testing this hypothesis. We noticed that the proposed theoretical mechanism only works when the lumen of the organoid is initially swollen, i.e. when fluid pressure exerts tension on the epithelium ($v > 1$ in the phase diagram Fig. 2A). We thus inferred the mechanical state of the lumen by experimentally measuring its volume before and after inducing epithelial breakage with 3D localised laser ablation of the tissue: if the lumen is initially swollen, tissue breakage will then let luminal fluid flow out, and vice versa. Importantly, these measurements (Fig. 2E and S1E) confirmed that bulged organoids are in a swollen state ($v \approx 2.77 \pm 1.27$, mean±SD) while budded samples are slightly shrunk ($v \approx 0.71 \pm 0.18$) – consistent with the 20-80% decrease in lumen volume observed during normal organoid morphogenesis from bulged to budded shapes (Yang et al, 2021).

Importantly however, with this experimentally inferred value of lumen volume, bistability is predicted to occur in a very limited parameter range of tensions $\sigma_c$ (range of 0.14 – 0.15, Fig. 2A), which would require extreme fine-tuning of cellular tensions in organoids. Given natural variabilities in Myosin and tension levels (Yang et al, 2021), this scenario would fail to ensure the robustness of crypt morphogenesis. Furthermore, given previous findings that Myosin intensity ratios can serve as a good proxy for tension ratios (Salbreux et al., 2012; Chugh and Paluch, 2018; Streichan et al., 2018), we quantified the intensities of the fluorescent reporter for the force-generating Non-Muscle Myosin II isoform (Myh9-GFP) in crypt apical and basal surfaces to calculate the in-plane to lateral tension ratio $\frac{\Gamma_a+\Gamma_b}{\Gamma_l}$ and in-plane tension ratio $\alpha$ (Fig. S6 and S7), as well as estimate the "tension asymmetry" in crypt apical and basal tensions, defined as $\epsilon = \frac{\Gamma_a-\Gamma_b}{\Gamma_a+\Gamma_b}$. Importantly, we found that tension asymmetries $\epsilon$ in budded crypts are below the theoretical threshold required for organoids to stay within the budded or bistable region of the phase diagram, and thus robustly remain budded, upon lumen inflation (Fig. 2F). This discrepancy with our experimental data (Fig. 1D) indicates that although the model can qualitatively resolve the paradox of distinct lumen inflation effects at different time points, other mechanisms must be at play to ensure that organoids robustly remain in the bistable region of the phase diagram (rather than re-entering the "bulged crypt" region) upon lumen inflation.

**Morphological bistability with mechano-sensitivity of crypts**

So far, we have assumed that lumen volume changes and actomyosin tensions are fully independent parameters. However, the volume and pressure of the luminal fluid have direct impact on the geometry and stresses of the crypt tissue, which could feedback on cellular surface tension via a variety of mechano-sensitive mechanisms, as described in other model systems (Desprat et al., 2008; Pouille et al., 2009; Fernandez-Gonzalez et al., 2009; Okuda et al., 2018). Furthermore, stem cells in crypts were found to be mechanosensitive as evidenced by the upregulation of Piezo1 and loss of stemness, upon organoid inflation (Tallapragada et al., 2021). Recent *in vitro* experiments on substrates of well-defined geometries have also revealed that intestinal crypt formation can be biased by substrate curvature (Nikolaev et al., 2020; Gjorevski et al., 2022; Pentinmikko et al., 2022), in line with a growing body of evidence of mechanosensation from tissue curvature (Mobasseri et al., 2019; Luciano et al., 2021; Tomba et al., 2022). Based on these findings, we considered two generic mechanisms of mechano-transduction: (i) stress-dependent feedback where the mechanical stresses in the crypt modulates actomyosin tensions; or (ii) curvature/geometry-dependent feedback, where the curvature of the crypt cells influences actomyosin tensions. Given that both models give rise to qualitatively similar theoretical predictions (Fig. S4-5, see SI Note for details), we concentrate here on curvature-dependent feedback, as curvature is easier to measure experimentally than stress (Fig. 3A, see SI Note for the alternative case). More specifically, we consider an equation $\sigma_c = \sigma \left(\frac{R_c}{\tilde{R}_0}\right)^{-n}$, where the crypt differential tension $\sigma_c$ depends both on an intrinsic value $\sigma$ (set by stem cell fate, Yang et al., 2021) and on the crypt radius of curvature $R_c$ (normalized by a reference value prior to morphogenesis $\tilde{R}_0$), with $n$ a coefficient quantifying the strength of the coupling.

Interestingly, such feedback functions distinctly in bulged vs budded crypts (Fig. 3B and S4): decreasing lumen volume in bulged organoids (as occurs in normal organoid development) causes an increase in crypt differential tension, because it affects crypt geometry. Thus, lumen deflation critically contributes to crypt morphogenesis, both by its direct mechanical impact (decreasing monolayer tension) and indirect mechanosensitive consequence (increasing actomyosin differential tension $\sigma_c$). However, once differential tension is sufficiently high that the crypt becomes budded, as previously discussed (Fig. 2A-C), the system is mechanically "trapped" in the budded shape as it becomes energetically more favourable to deform villus, rather than crypt, cells upon inflation.

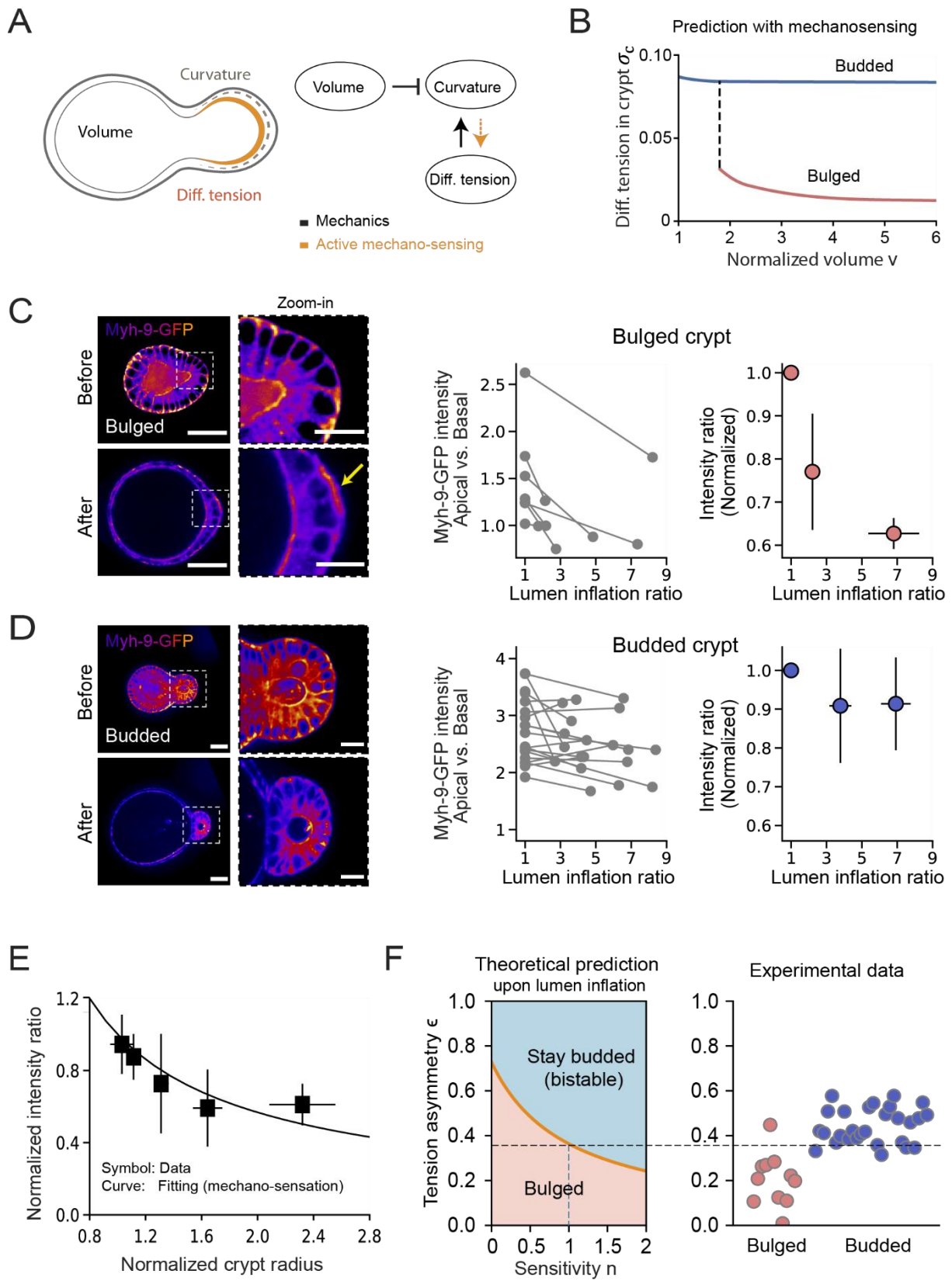

**Figure 3: Morphological bistability with mechano-sensitivity of crypts. A**. Schematic of the feedback mechanism that involves both mechanical forces and mechano-sensitivity of crypt cells: lumen volume and apical to basal tension difference affect the curvature of the crypt due

to passive force balance, while active mechano-sensing can result in the geometry/mechanics of the crypt feedbacking on apical/basal tensions. **B**. Theoretical prediction of the dependence of crypt differential tension on lumen volume, in both bulged and budded organoids, when assuming mechano-sensation: in bulged organoids, inflation results in crypt opening which negatively regulates tensions by mechano-sensing, while budded organoids have structurally stable crypts that do not open upon inflation, which thus does not trigger an active response. **C, D**. Left, Myh9–GFP distribution on crypt surfaces, before and after lumen inflation. After lumen inflation, bulged crypts show basal actomyosin relocation (yellow arrow). Right, apical to basal Myh9–GFP intensity ratio in bulged (C, $N = 7$) and budded (D, $N = 19$) crypt cells. **E**. Experimental data (symbols) and fitting (line) of mechano-sensitivity, both apical to basal Myh9–GFP intensity ratio and crypt radius after lumen inflation are normalized by their values before inflation. Both bulged ($N = 24$) and budded samples ($N = 28$) are included. The fitting curve is $y = x^{-b}$. **F**. Influence of sensitivity factor $n$ on the predicted threshold for bistability, and comparison with experimental data of bulged ($N = 11$) and budded ($N = 28$) samples for estimated value of $n = 1.0$ (inferred from the best-fit value in E, see main text and SI note for details). Scale bars in images: 50 µm (organoid) and 20µm (magnification). Error bars represent SD.

**Lumen volume changes affect actomyosin localization**

To test these predictions, we first experimentally validated the key modelling assumption that differential tension is dependent on crypt geometry and lumen volume. We thus increased lumen volume in early bulged or late budded organoids, by treatment with prostaglandin E2 (PGE). We had previously shown that this results in crypt opening in bulged organoids, but no change in budded crypt geometry (Yang et al., 2021). Importantly, we found that these volume perturbations resulted in very different consequences on the levels and localization of Myosin in bulged vs budded crypts. While bulged crypts displayed lowered apical to basal Myosin ratio upon volume inflation in a dose-dependent manner (Fig. 3C), budded crypts did not show a consistent change (Fig. 3D). This argues that the changes in actomyosin localization in bulged crypts are not a secondary effect of osmotic changes upon inflation, but are instead linked to changes in crypt geometry. Indeed, pooling all bulged and budded inflation data together, we found a consistent trend clearly linking the relative actomyosin intensity to crypt curvature (Fig. 3E). We then used this general relationship to quantitatively parametrize the mechano-sensitive feedback between apico-basal tensions and geometry in the model (i.e. $\bar{\epsilon} = \bar{R}_c^{-n}$), leading to a fit for the mechano-sensitivity factor of $n = 1.0 \pm 0.5$ (see SI Note for details).

Upon incorporation of this mechano-sensitive coupling parameter, the theory predicted much broader regions of shape bistability (Fig. S4C-D, for instance at $v = 3$, bistability occurs when $\sigma$ is between 0.02 – 0.04, a range which was an order of magnitude broader than without mechano-sensing), which allows for robust and irreversible budding of crypts with various intrinsic tension. Quantitatively, mechano-sensing lowered the differential tension threshold for bistability (i.e. the boundary between bulged and budded phases). As a key consequence, we found that the tension asymmetry $\epsilon$ previously inferred from Myh-9-GFP in budded crypts, is now above the theoretical threshold for bistability, enabling budded organoids to robustly remain budded upon arbitrarily large lumen inflation (Fig. 3F, 4A-B). As an additional quantitative control, we also found that the threshold was still above the tension asymmetry $\epsilon$ from budged crypts, supporting that the experimental observation that bulged crypt require lumen volume decrease to bud (Fig. 3F, 4A-B).

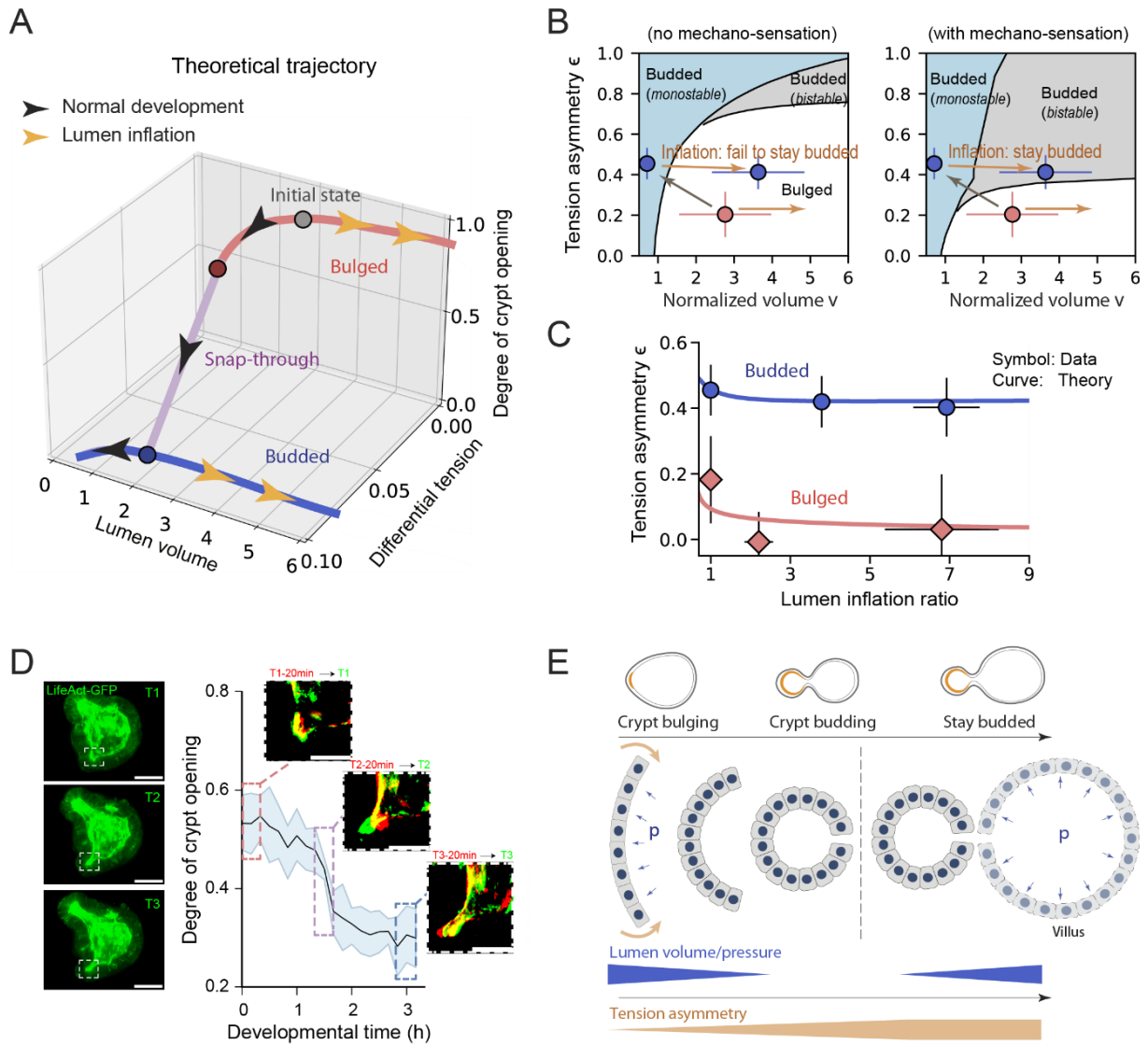

**Figure 4: Theoretical predictions of mechano-chemical bistability for crypt morphogenesis. A**. Theoretical bifurcation trajectories showing the evolution of crypt morphology and differential tension with lumen volume, specified as normal morphogenesis (black arrows) and lumen inflation of both bulged and budded organoids (orange arrows). **B**. Comparison between experimental data (red dot: bulged samples $N = 11$, blue dots: budded samples $N = 28$) and predicted phase diagrams of crypt morphology as a function of tension asymmetry $\epsilon$ and normalized volume $v$. Tension asymmetry of budded organoids agrees with the prediction considering crypt mechano-sensation (right panel) rather than that without mechano-sensation (left panel, see also Fig. 2F). **C**. Predicted evolution for tension asymmetry data of both bulged ($N = 7$) and budded ($N = 19$) samples with the theoretical model, where the intrinsic differential tension $\sigma$ is the only free parameter fitted as 0.02 (see SI Note for details). **D**. Evolution of crypt shape with developmental time ($N = 5$), showing rapid changes in shape at a critical morphological point, as expected from bistability. For different developmental stages, changes of crypt profiles in time interval 20 min (red, start of the interval; green, end of the interval) are shown. **E**. Schematic of crypt morphogenesis driven by luminal pressure, and involved morphological bistability feature arising from mechano-sensation. Scale bars in images: 50 μm. Error bars represent SD.

**Quantitative morphogenetic trajectories of intestinal organoids**

To summarize these findings, we derived the predicted trajectory of organoids with mechano-sensing (Fig. 4A), which can reproduce the bistable hysteresis on both crypt morphology and crypt cellular tensions (Fig. 4B). Lumen shrinkage is critical during normal crypt morphogenesis, by both decreasing the cost associated with tissue bending (passive mechanical effect), and increasing differential tension (active mechano-sensing effect). In contrast, subsequent volume inflation in budded organoids does not cause the system to transition back to open crypt morphologies and crypt cellular tensions are maintained. More quantitatively, the model can simultaneously fit cell tension data from both bulged and budded samples (Fig. 4C), with the single fitting parameter (intrinsic differential tension in the crypt) $\sigma = 0.02$. Finally, a non-trivial prediction from this model is that crypt morphologies should first vary relatively little with increasing apical tension and decreasing volume, before abruptly changing at the critical transition point. Thus, although crypt apical tension and lumen volume change gradually (Yang et al., 2021), this would predict that the transition to a budded crypt should be fast as the organoid switches from one stable minimum of its energy landscape to another (snap-through bifurcation). Interestingly, plotting the crypt opening angles for different organoids as a function of time revealed a phase of abrupt closure (within 20min, Fig. 4D), compared to a total time of around 24h for full organoid morphogenesis (Fig. 1A and Yang et al., 2021).

**Discussion**

In this work, we have combined a minimal description of three-dimensional epithelial monolayer shape, together with mechano-sensitive couplings. We constrain this theory based on detailed morphometric measurements of cell shape in intestinal organoids as well as live reporters of Myosin activity upon mechanical perturbations, namely lumen inflation. We show that a bistable region in organoid phase-space generically arises, which can explain a number of experimental features on how lumen volume changes can mediate different responses at different morphogenetic timepoints (Fig. 1C, 4A). Indeed, the presence of bistability confers strong hysteretic behaviour to the system, so that lumen volume changes can both be a key control parameter during the regular process of organoid morphogenesis, as well as be rendered irrelevant once budded crypt morphogenesis is completed (Fig. 4E). This bears conceptual similarities to the role of pre-patterned vs self-organized/self-reinforcing cues in other developmental systems, such as the mechano-chemical interplay of embryo polarization in *C. elegans* (Sailer et al, 2015, Gross et al., 2019). This might therefore be a general mechanism that could ensure the robustness and irreversibility of morphogenesis in multicellular systems.

Our work also provides an example of the emerging role of fluid lumen pressure in controlling morphogenesis (Saias et al., 2015; Chan et al., 2019; Dumortier et al., 2019; Munjal et al., 2021). Compared to other types of biological forces, fluid pressure has the advantage of being intrinsically long-ranged. For instance, in intestinal morphogenesis, an outstanding question that remains is the nature of the coordination between the specification and maturation of different crypts, both in vivo and in vitro. In organoids, lumen volume changes would be expected to provide a global cue that is felt equally among all crypts of an organoid regardless of its size. Coupled to mechano-sensing, such a mechanism could allow for synchronization of crypt morphogenesis, especially as the earlier crypt fate specification events are rather asynchronous (Serra et al, 2019). Given that crypt fission was also shown to be dependent on lumen volume changes (Tallapragada et al., 2021), such mechanism could also act at multiple different time points during development. Importantly, even without fluid pressure, in vivo crypts can still be subjected to other external forces, arising for instance from the constriction of the smooth muscle layers (Shyer et al., 2013) or from the osmotic swelling of villus cells (Yang et al., 2021). These forces would also be a source of coordination by generating bistability. In the future, it will be important to understand, at the cellular and molecular level, the mechanisms of the relationship between osmotic forces, lumen volume, crypt geometry and actomyosin accumulation, as well as to test whether similar principles of bistability and robustness of morphogenesis hold in other developmental and organoid systems.

**Acknowledgements**

We thank all members of the Hannezo and Liberali groups for fruitful discussions, as well as Cornelia Schwayer, Gustavo Quintas, David Bruckner and Diana Pinheiro for reading the manuscript. This work received funding from the European Research Council (ERC) under the European Union's Horizon 2020 research and Innovation Programme Grant Agreement no. 758617 (to P.L.) and Swiss National Foundation (SNF) (POOP3_157531 to P.L.), also from the ERC under the European Union's Horizon 2020 Research and Innovation Program Grant Agreements 851288 (to E.H) and the Austrian Science Fund (FWF) (P 31639 to E.H.).

**Author contributions**

E.H., S.L.X., Q.Y. and P.L. conceived the project and designed the experiments. Q.Y. performed the experiments. Q.Y. and S.L.X. analysed the experimental data. S.L.X. designed the physical model and performed the simulations. S.L.X. wrote the first version of the manuscript, E.H., S.L.X., Q.Y. and P.L. reviewed and wrote the manuscript.

**Competing interests**

Authors declare no competing interests.

**Data and code availability**

Source data will be submitted for all the plots. All other data supporting the findings of this study are available from the corresponding authors on request. Codes used in the current study are from previous work and available on
https://github.com/fmi-basel/glib-nature_cell_biology2021-materials.git.

## Materials and Methods

### Animal work

All animal experiments were approved by the Basel Cantonal Veterinary Authorities and conducted in accordance with the Guide for Care and Use of Laboratory Animals. Male and female outbred mice from 7 weeks old onwards were used for generating organoid lines of C57BL/6 wild type and Lgr5-DTR-EGFP as reported previously (Yang et al., 2021). One male mouse at 10 weeks was used to generate organoid line of LifeAct-GFP. One male mouse at 10 weeks was used to generate organoid line of Myh-9-GFP.

Mouse lines used: C57BL/6 wild type (Charles River Laboratories), Lgr5-DTR-EGFP (Genentech, de Sauvage laboratory), LifeAct-GFP (T. Hiiragi laboratory, EMBL), Myh-9-GFP (Lennon-Duménil laboratory, Institut Curie).

Mice were kept in housing conditions with 12-hour light/12-hour dark cycle, 18-23 °C ambient temperature and 40-60% humidity.

### Organoid culture

Organoids were generated from isolated crypts of the murine small intestine as previously described (Yang et al., 2021). Organoids were kept in IntestiCult Organoid Growth Medium (STEMCELL Technologies) with 100 μg/ml Penicillin-Streptomycin for amplification and maintenance.

### Time course experiments of fixed organoid samples

The method was adapted from described before (Yang et al., 2021). Organoids were collected 5-7 days after passaging and digested with Tryple (Thermo Fisher Scientific) for 20 min at 37 °C. Dissociated cells were passed through a cell strainer with a pore size of 30 μm (Sysmex). Collected cells were mixed with Matrigel (Corning) in a ENR medium to Matrigel ratio of 1:1(15). In each well of a 96 well plate, 5μl droplets with 2500 cells were seeded. After 15 min of solidification at 37 °C, 100 μl of medium was overlaid. From day 0 to day 1, ENR was supplemented with 25% Wnt3a-conditioned medium (Wnt3a-CM), 10 μM Y-27632 (ROCK inhibitor, STEMCELL Technologies) and 3 μM of CHIR99021 (GSK3B inhibitor, STEMCELL Technologies, cat # 72054). From day 1 to 3 ENR was supplemented with 25% Wnt3a-CM and 10 μM Y-27632. From day 3 to 5, only ENR was added to the cells. Wnt3a-CM was produced in-house by Wnt3a L-cells (kind gift from Novartis).

### Compound treatments

Compound treatments were tested in dilutions series of various concentrations from 1mM to 5nM in previous study (Yang et al., 2021).

Single cells derived from LifeAct-GFP organoids were plated in a 96-well plate chamber and exposed to 0.5 µM Prostaglandin E2 (PGE, kind gift from Novartis) or 0.5µM DMSO (SIGMA-ALDRICH, cat # D8418) diluted in ENR medium, from 72 hours for 24hrs (DMSO, Fig. 1A), or from 96 hours for 2 hours (PGE, Fig. 1B), or from 72 hours for 10 hrs (PGE, Fig.1C).

Single cells derived from Myh-9-GFP organoids were plated in an ibidi 8µ plate chamber and culture in time-course medium till 84 hours or 96 hours for laser nanosurgery (Fig. 2E) or PGE treatment (Fig. 3C and D).

Single cells derived from Lgr-5-DTR-GFP organoids were plated in a 96-well plate chamber and culture in time-course medium until fixation at 72 or 84 or 96 hours (Fig. S1).

Single cells derived from organoids C57BL/6 wild type were plated in a 96-well plate and treated with 0.5 µM Prostaglandin E2 (PGE, kind gift from Novartis) or 0.5 µM DMSO in ENR medium, from 72 hours until fixation at 74hours or 84hours or 96 hours, or from 96 hours until fixation at 98 hours (Fig. 1D).

**Organoid immunostaining and imaging**

The method was adapted from descripted before (Yang et al., 2021). Primary and secondary antibodies were diluted in blocking buffer and applied. Cell nuclei were stained with 20 µg/ml DAPI (4',6-Diamidino-2-Phenylindole, Invitrogen) in PBS for 5 min at room temperature. Cells were stained with 1 µg/ml of Alexa Fluor® 647 carboxylic acid succinimidyl ester (CellTrace, Invitrogen) in carbonate buffer (1.95 ml of 0.5 M NaHCO3, 50 µl of 0.5 M Na2CO3, both from Sigma-Aldrich, in 8 ml of water for 10 ml of buffer).

High-throughput imaging was done with an automated spinning disk microscope from Yokogawa (CellVoyager 7000S), with an enhanced CSU-W1 spinning disk (Microlens-enhanced dual Nipkow disk confocal scanner), a 40x (NA = 0.95) Olympus objective, and a Neo sCMOS camera (Andor, 2,560 × 2,160 pixels). For imaging, an intelligent imaging approach was used in the Yokogawa CV7000 (Search First module of Wako software 1.0) as described before (Yang et al., 2021). Z-planes spanning a range up to 90 µm and 2 or 3 µm z-steps were acquired.

Confocal imaging of fixed samples was performed using Nikon Ti2-E Eclipse Inverted motorized stand with Yokogawa CSU W1 Dual camera (CAM1 SN: X-11424; CAM2 SN:11736) T2 spinning disk confocal scanning unit, CFI P-Fluor 40x oil/1.4 objective and Visview 4.4.0.9 software. Laser lines used are Toptica iBeam Smart 405/488/639 nm and Cobolt Jive 561 nm. Laser power and digital gain settings were unchanged within a given session to permit direct comparison of expression levels among organoids stained in the same experiment. Image stacks were acquired with slice thickness of 2µm or less.

Confocal imaging of live Myh-9-GFP samples for two time-points before and after PGE-

inflation was performed using the same microscope and imaging sitting as confocal imaging of fixed samples. Before PGE-treatment, one to five organoids from one well were quickly imaged within 10 minutes. Then PGE was added into the culture medium of the same well and put back to 37°C tissue culture. After 20 to 30 minutes, organoid lumens are sufficiently inflated. Imaging was taken from the same organoids for the time point after inflation.

**Time-course image analysis**

Organoid segmentation in maximum intensity projections (MIPs) was adapted from before (Yang et al., 2021). For each acquired confocal z-stack field, MIPs were generated. All MIP fields of a well were stitched together to obtain MIP well overviews for each channel. The high resolution well overviews were used for organoid segmentation and feature extraction. From the segmented MIPs, we measure and calculate the features of each individual organoid.

**Light-sheet microscopy**

Light-sheet microscopy was conducted by using LS1 Live light sheet microscope system (Viventis) or a similar customized microscope system as described before (Yang et al., 2021). Sample mounting was performed as described previously (Yang et al., 2021). For organoids imaging, LifeAct-GFP organoids were collected and digested with TrypLE (Thermo Fisher Scientific) for 20 min at 37 °C. GFP positive cells were sorted by FACS and collected in medium containing advanced DMEM/F-12 with 15 mM HEPES (STEM CELL Technologies) supplemented with 100 μg/ml Penicillin-Streptomycin, 1×Glutamax (Thermo Fisher Scientific), 1×B27 (Thermo Fisher Scientific), 1xN2 (Thermo Fisher Scientific), 1mM N-acetylcysteine (Sigma), 500ng/ml R-Spondin (kind gift from Novartis), 100 ng/ml Noggin (PeproTech) and 100 ng/ml murine EGF (R&D Systems). 2000 cells were then embedded in 5μl drop of Matrigel/medium in 50/50 ratio. Drops were placed in the imaging chamber and incubated for 20 min before being covered with 1ml of medium. For the first three days, medium was supplemented with 20% Wnt3a-CM and 10 μM Y-27632 (ROCK inhibitor, STEMCELL Technologies). For the first day, in addition, 3 μM of CHIR99021 (STEMCELL Technologies) were supplemented. After more than 2 days culture in a cell culture incubator the imaging chamber was transferred to the microscope kept at 37 °C and 5% CO2. Different organoids were selected as starting positions and imaged every 10 min. A volume of 150-200μm was acquired with a Z spacing of 2μm between slices. Medium was exchanged manually under the microscopy every half day.

**Data analysis**

Tissue opening angles $\theta_i$ (see Fig. 2A for schematic, with $i = c, v$ representing crypt and villus tissue), which related to the degree of crypt opening (defined as $\theta_c/(\pi - \theta_v)$), can be inferred from the coordinate information of characteristic points along the mid-plane of tissue, by approximating the tissue as a spherical cap. For each region (crypt or villus) in each sample, extract three points from the organoid image by using the multi-point tool of Fiji (National Institutes of Health): two at the boundary (crypt/villus), denoted as $(x_1, y_1)$ and $(x_3, y_3)$, and

one in the middle $(x_2, y_2)$ (see Fig. S1D for schematic). The opening angle can be calculated as

$\theta = \pi - \arccos \frac{(x_1-x)(x_2-x)+(y_1-y)(y_2-y)}{\sqrt{[(x_1-x)^2+(y_1-y)^2]\cdot[(x_2-x)^2+(y_2-y)^2]}}$, which involves the coordinate of the center of the spherical tissue: $x = \frac{B}{2A}$, $y = -\frac{C}{2A}$, with $A = x_1(y_2 - y_3) + x_2(y_3 - y_1) + x_3(y_1 - y_2)$, $B = (x_1^2 + y_1^2)(y_2 - y_3) + (x_2^2 + y_2^2)(y_3 - y_1) + (x_3^2 + y_3^2)(y_1 - y_2)$, $C = (x_1^2 + y_1^2)(x_2 - x_3) + (x_2^2 + y_2^2)(x_3 - x_1) + (x_3^2 + y_3^2)(x_1 - x_2)$.

Morphometric parameters, including cell height $h_i$, cell width $d_i$ and tissue radius $R_i$, were extracted from images of intestinal organoids by using Fiji. Cell height (i.e. epithelial thickness) $h_i$ and cell width $d_i$ of each region (crypt or villus) were measured by the line tool of Fiji in at least three random positions and determined as the average value of these measurements. Tissue radius $R_i$ is defined as the average value of the apical and basal radii (see SI note for details), and was quantified by the oval tool of Fiji.

Myh-9-GFP intensity ratios in single cells were measured using Fiji as described before (Yang et al., 2021).

Lumen volume calculation of light sheet data was proceeded as previously (Yang et al., 2021).

**Lumen breakage by laser nanosurgery**

The method was adapted from previous studies (Yang et al., 2021). In brief, images are captured by a LSM710 scanning confocal microscopy using ZEN Black software (the only version). The microscope is equipped with an incubation chamber to keep the sample at 37°C and to provide 5% of $CO2$. Organoids were embedded in Matrigel and cultured in ibidi 8-well plates. Cutting was performed using a wavelength of 850nm with a Chameleon Ultrall Laser. All cuttings were performed in the villus region of organoid. The cutting region (yellow dashed lines in Fig. 2E) was set as a rectangle of 0.9 μm x 50 μm, and the activation time was calculated by the scan speed of 1.25ms/pixel. Organoids were imaged with a Plan-Neofluar 40x/0.9 Imm Korr Ph3 Objective lens before and after cutting.

**Statistics and reproducibility**

All statistical analysis was performed using the python library SciPy. All Box-plot elements show 25% (Q1, upper bounds), 50% (median, black lines within the boxes) and 75% (Q3, lower bounds) quartiles, and whiskers denote 1.5× the interquartile range (maxima: Q3 + 1.5× (Q3-Q1); minima: Q1-1.5× (Q3-Q1)) with outliers (rhombuses). In compound time-course experiments, we assumed a minimum of around 50 organoids at Day4 would be sufficient to recognize differences between control and perturbations based upon historical experiments. Sample size was determined based on previous related studies in the field (Serra et al., 2019).

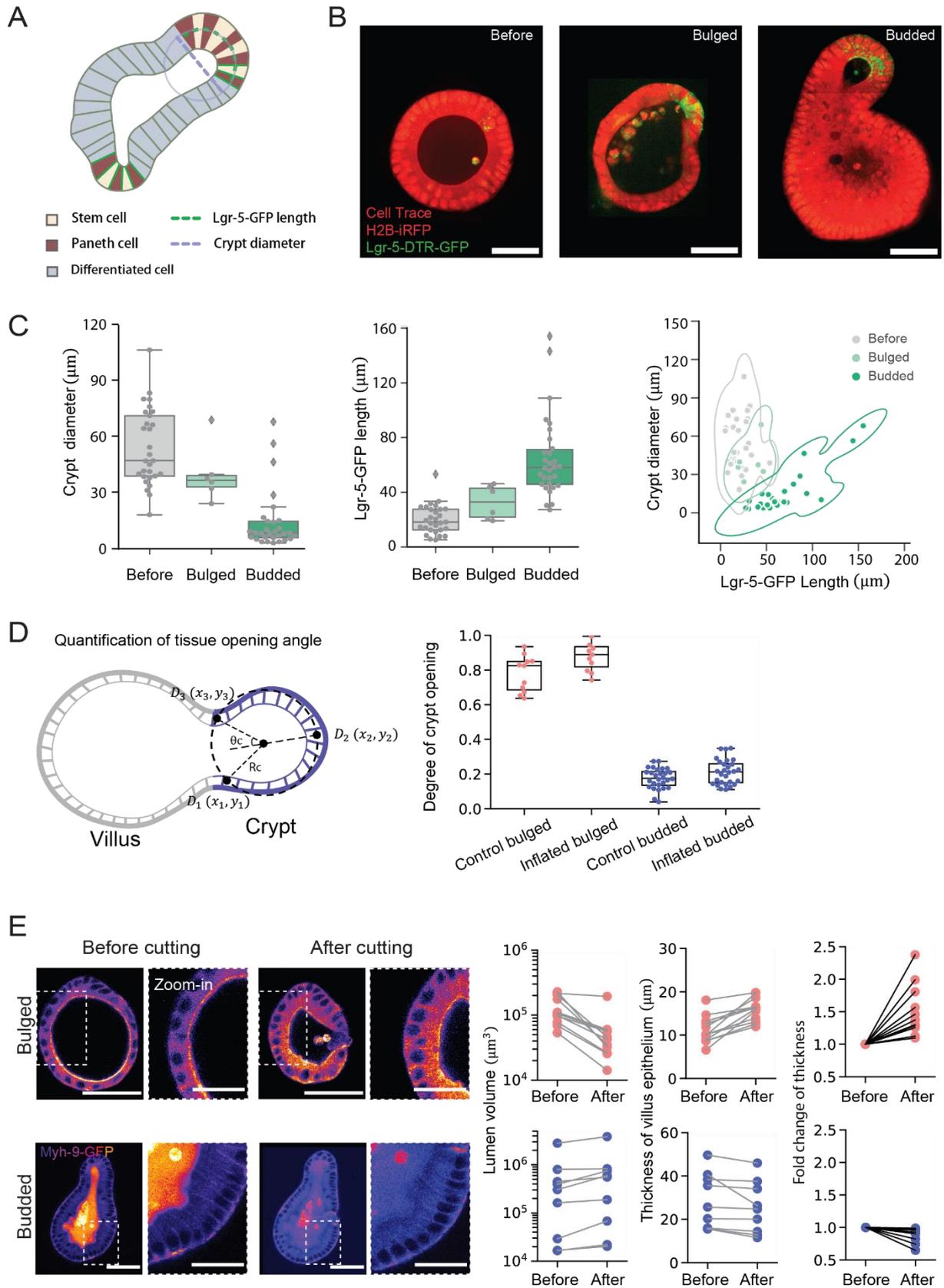

**Figure S1: Quantification of tissue morphology and lumen volume. A, B**. Cartoon representation of the quantified morphometric parameters (A) and representative time-course

images of crypt morphogenesis (B). Cell trace indicates all cells, H2B-iRFP marks cell nuclei and Lgr-5-DTR-GFP marks intestinal stem cells. **C**. Crypt diameter (left), Lgr-5-GFP length (middle), and crypt diameter against Lgr-5-GFP length (right) for organoids at different stages of morphogenesis (before: $N = 29$; bulged: $N = 6$; budded: $N = 30$). Crypt curvature and Lgr-5-GFP positive region increase during crypt morphogenesis. **D**. Quantification of crypt opening degree. Left, schematic to show data extraction for quantifying tissue opening angle (see Methods for details). Right, degree of crypt opening before and after lumen inflation, in both bulged ($N = 11$) and budded ($N = 28$) organoids. **E.** Quantification of lumen volumes before and after villus tissue breakage, in both bulged (top panels, $N = 11$) and budded (bottom panels, $N = 9$) organoids. Epithelial thickness in villus (absolute values and fold changes) can also infer the lumen volume changes: bulged samples show increased thickness of villus tissue after tissue breakage (top), indicating lumen volume decrease that is consistent with the volume measurement. Budded ones show the reverse trend (bottom). Scale bars: 50 µm (organoid) and 25µm (magnification).

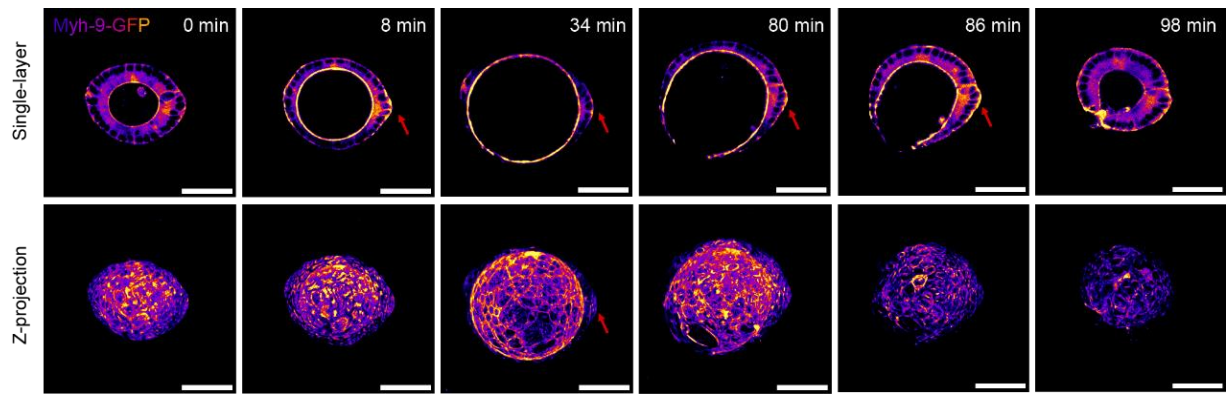

**Figure S2: Time-lapse recordings of basal crypt actomyosin relocation during lumen inflation.** Red arrows point to the crypt basal surface with increasing Myh-9-GFP intensity. Scale bars: 50 µm.

## A

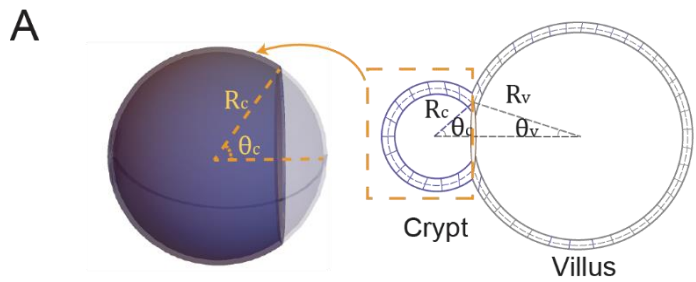

Tissue as spherical cap:

Volume: $V_i = \pi R_i^3 (2 + 3\cos\theta_i - \cos^3\theta_i)/3$
$(i = c, v)$

Cell number: $N_i/N_i' = (1 + \cos\theta_i)/2$

tissue (spherical cap) ↗  ↖ whole sphere

## B

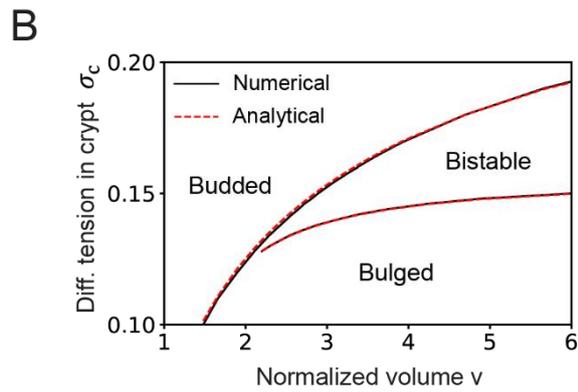

## C

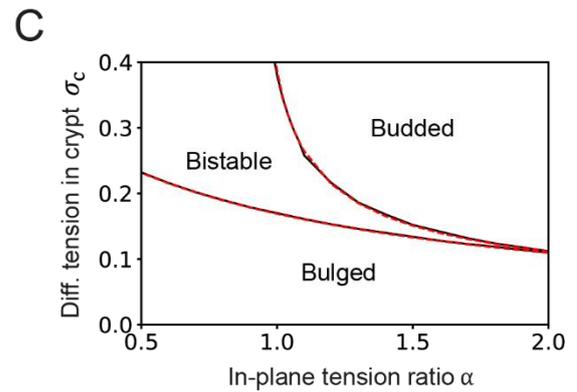

## D

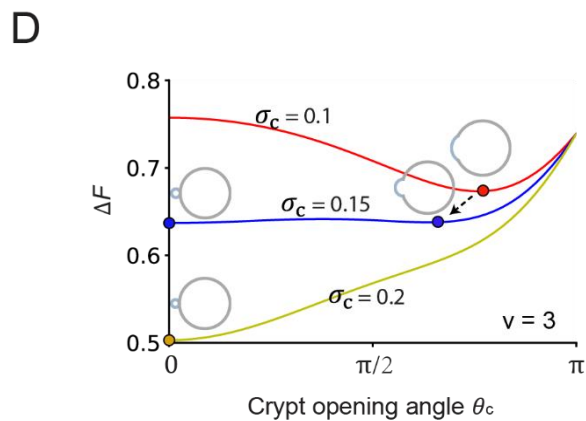

## E

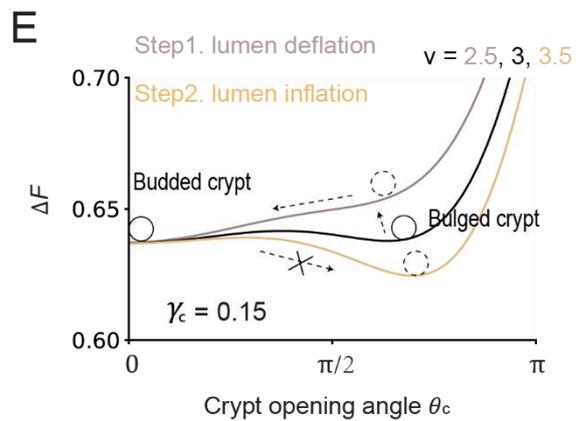

## F

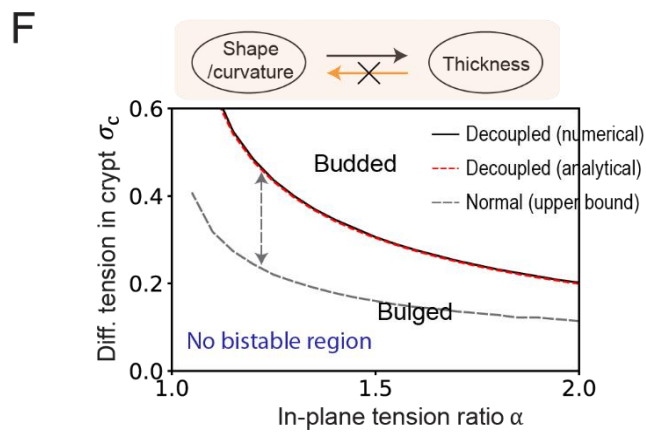

**Figure S3: Detailed theoretical analysis of crypt morphological bistability without mechano-sensation.** **A**. Three-dimensional geometry of intestinal organoid, and basic geometric relations used to establish the model (see SI Note for details). **B, C**. Comparison between numerical (from the full model) and analytical (from the simplified model) results of phase boundaries in $\sigma_c - v$ diagram ($\alpha = 1.2$) (B) and $\sigma_c - \alpha$ diagram ($v = 10$) (C), showing high consistency. **D, E**. Mechanical energy profile $\Delta F$ as a function of crypt opening angle $\theta_c$ for various values of differential tension $\sigma_c$ (D) and lumen volume $v$ (E). Arrows indicates the evolution of crypt shape with relevant parameters. **F**. Crypt morphology in $\sigma_c - \alpha$ diagram after decoupling crypt curvature and thickness, and comparison with the original model. Compared to panel C, this shows an absence of bistable region, as well as higher threshold for budding.

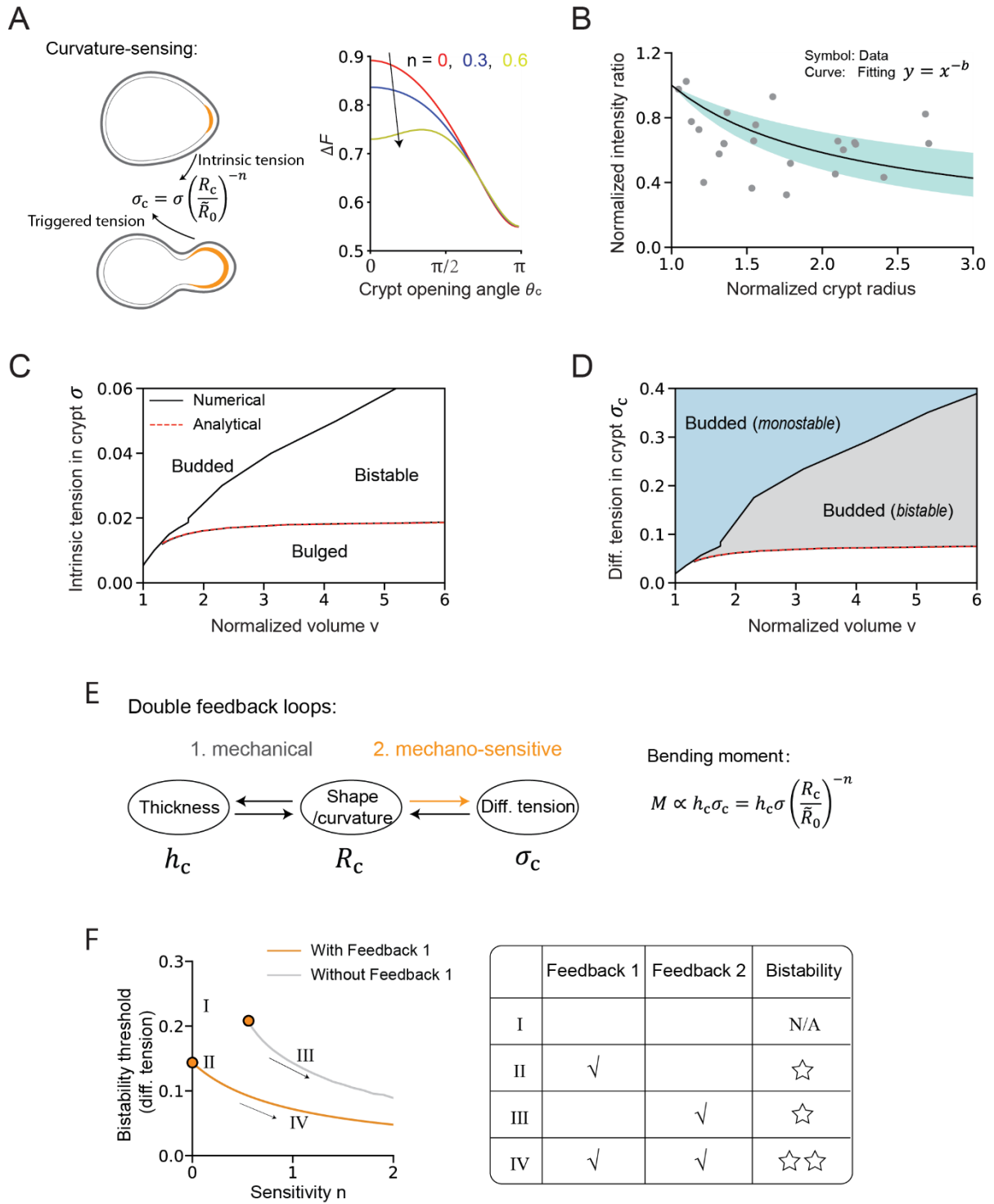

**Figure S4: Detailed analysis of crypt morphological bistability with curvature-sensitivity.** We set crypt size $\varphi = 0.1$ and in-plane contraction ratio $\alpha = 1.2$ for all theoretical predictions (based on quantifications in Fig. S7 for $\alpha$ and $\varphi$, and see SI Note for details). **A.** Left, schematic of the curvature-sensing mechanism, where the total differential tension $\sigma_c$ is the product of an intrinsic value $\sigma$ (arising from stem cell fate, Yang et al, 2021) as well as a contribution inversely proportional to the crypt radius of curvature $R_c$ with sensitivity $n$. Right, influence of curvature-sensitivity ($n$) on the energy landscape ($\sigma = 0.05$, infinite

volume), showing that it gives rise to a second minimum. **B**. Fitting of curvature-sensitivity. Apical to basal Myh9–GFP intensity ratio and crypt radius after lumen inflation are normalized by their values before inflation. Best fit of data (bulged samples, N = 24) and 95% confidence interval are respectively shown as solid line and shaded region. **C, D**. Crypt morphology with curvature-sensation ($n = 1$) in $\sigma - v$ diagram (C) and corresponding $\sigma_c - v$ diagram showing an enhanced region of bistability (D, compare to Fig. 2A and S3B). **E**. Double feedback loops in crypt morphogenesis. Feedback 1 is the mechanical coupling between epithelial thickness and curvature (see Fig. 2D for schematic), and Feedback 2 is the coupling between crypt tension and deformation (with mechano-sensation, see panel A for schematic). **F**. Evaluation of specific feedback loops that enable crypt bistability. Left, bistability threshold of differential tension $\sigma_c$ without/with the purely mechanical Feedback 1 as well as its dependence on Feedback 2 (mechano-sensation with sensitivity $n$). Right, summary of the effects of two feedback loops on the size of the bistable region in phase diagram. Error bars represent SD.

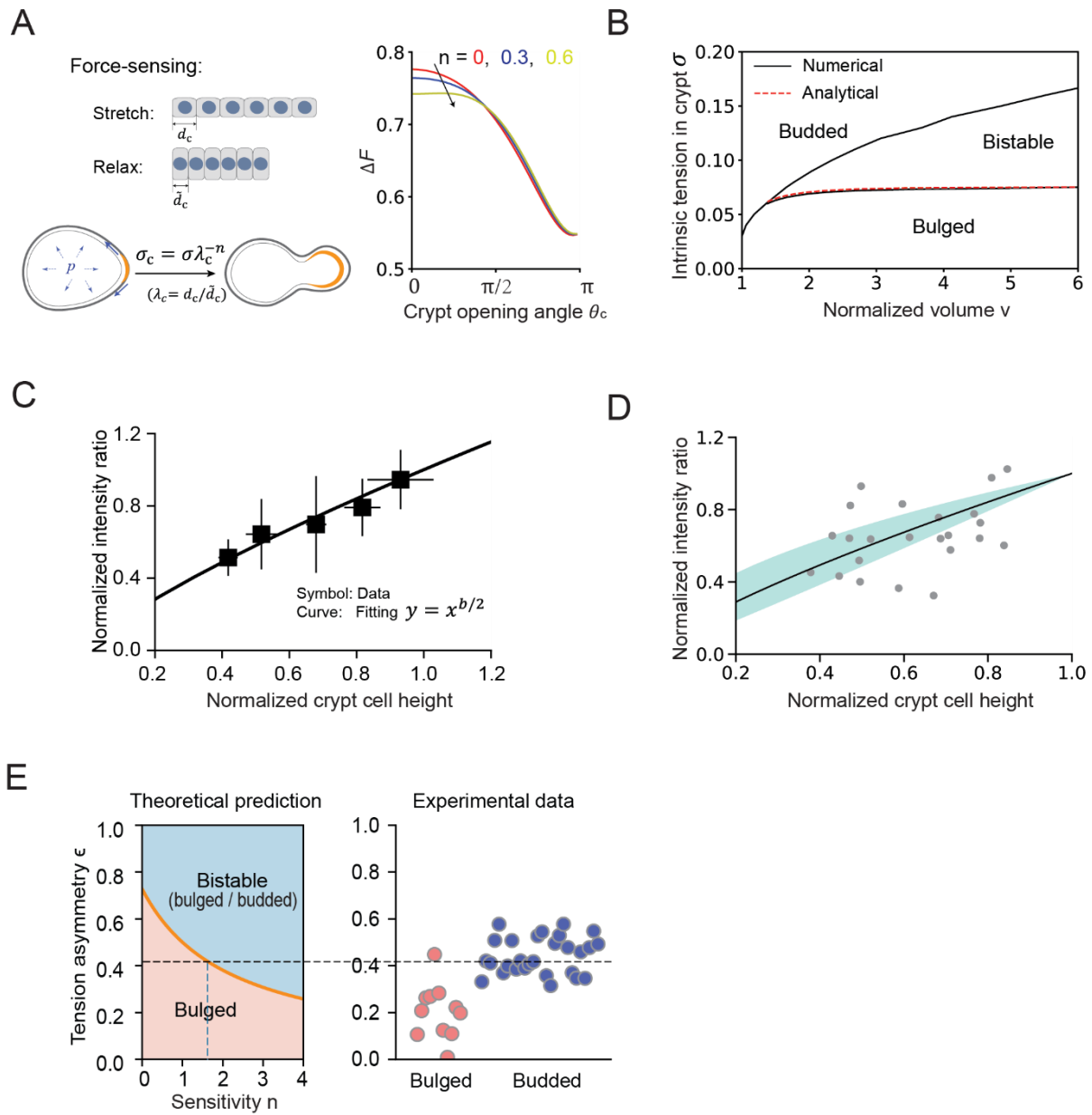

**Figure S5: Crypt morphological bistability with force-sensitivity. A**. Left, schematic of force-sensing mechanism, with crypt cell in-plane stretch ratio $\lambda_c$ defined. Right, influence of force-sensitivity ($n$) on the energy landscape ($\sigma = 0.1$, infinite volume), showing qualitatively similar features as curvature-sensing. **B**. Phase diagram of crypt morphology with force-sensation. **C, D**. Fitting of force-sensitivity. Apical to basal Myh9–GFP intensity ratio and crypt cell height (or thickness) after lumen inflation are normalized by their values before inflation. The fitting curve is $y = x^{b/2}$, with the fitting parameter $b$ used to infer force-sensitivity $n$. Fittings the mean values (squares, panel C) and raw data (dots, panel D) give similar best-fit values (see SI Note for details). The shaded region in D is 95% confidential interval. **E**. Predicted influence of sensitivity factor $n$ on the bistability threshold of tension asymmetry $\epsilon$, and comparison with experimental data of bulged ($N = 11$) and budded ($N = 28$) samples for the best-fit value $n = 1.7$ from panel C. Error bars represent SD.

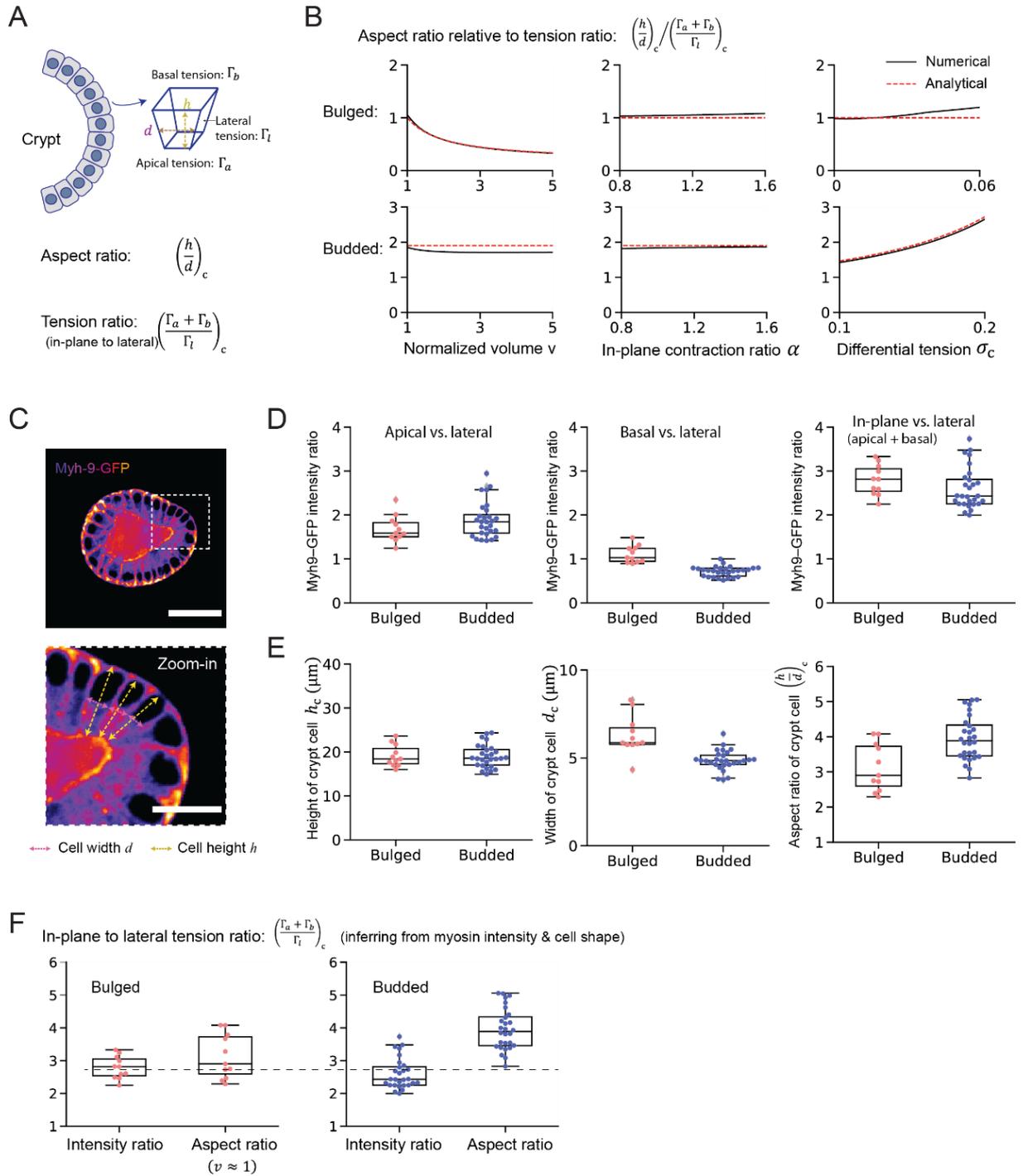

**Figure S6: Link between mechanical and morphometric parameters. A**. Definitions of crypt cellular aspect ratio $\left(\frac{h}{d}\right)_c$ and in-plane to lateral tension ratio $\left(\frac{\Gamma_a+\Gamma_b}{\Gamma_l}\right)_c$. **B**. Sensitivity analysis for how model parameters affect the correlation between aspect ratio $\left(\frac{h}{d}\right)_c$ and in-plane to lateral tension ratio $\left(\frac{\Gamma_a+\Gamma_b}{\Gamma_l}\right)_c$ in both bulged and budded crypts: the dependence of the normalized value $\left(\frac{h}{d}\right)_c / \left(\frac{\Gamma_a + \Gamma_b}{\Gamma_l}\right)_c$ on normalized volume $v$ (left), in-plane contraction ratio $\alpha$ (middle) and differential tension $\sigma_c$ (right). We set $v = 1, \alpha = 1.2$ and $\sigma_c = 0.03$ (bulged) or 0.15 (budded) unless the parameter is varying. **C**. Representative organoid image with Myh-9-GFP staining and extraction of cellular height $h$ and width $d$. **D**. Quantification of Myh-9-GFP intensity ratio: apical, basal, and in-plane (sum of apical and basal) intensities are normalized by lateral intensity in each crypt (bulged: $N = 11$, budded: $N = 28$). **E**. Quantification of cellular height $h_c$, width $d_c$ and aspect ratio $\left(\frac{h}{d}\right)_c$ in both bulged and budded crypts. Bulged samples are those after lumen breakage (i.e. $v \approx 1$, also shown in Fig. 2E), and in this state, cellular tension ratio and aspect ratio are expected to share the same value based on sensitive analysis B. **F**. Inference of in-plane to lateral tension ratio $\left(\frac{\Gamma_a + \Gamma_b}{\Gamma_l}\right)_c$ from corresponding Myh-9-GFP intensity ratio (i.e. right panel in D) and cellular aspect ratio (based on sensitive analysis B, tension ratio can also be inferred from aspect ratio in panel E). Scale bars: 50 μm (organoid) and 20μm (magnification).

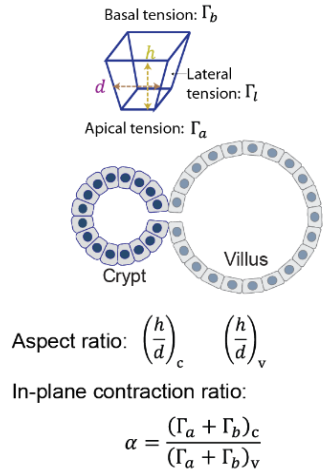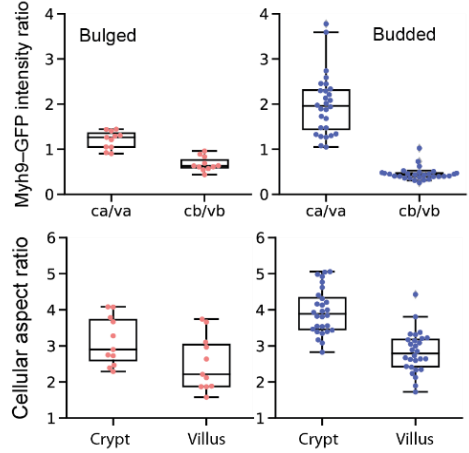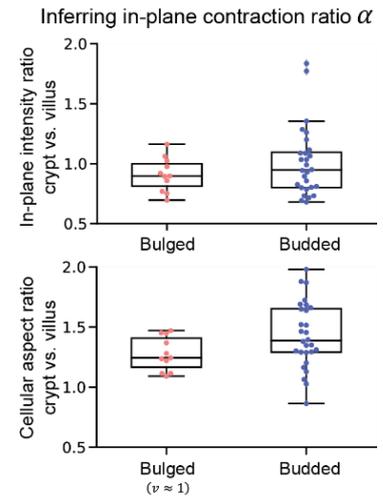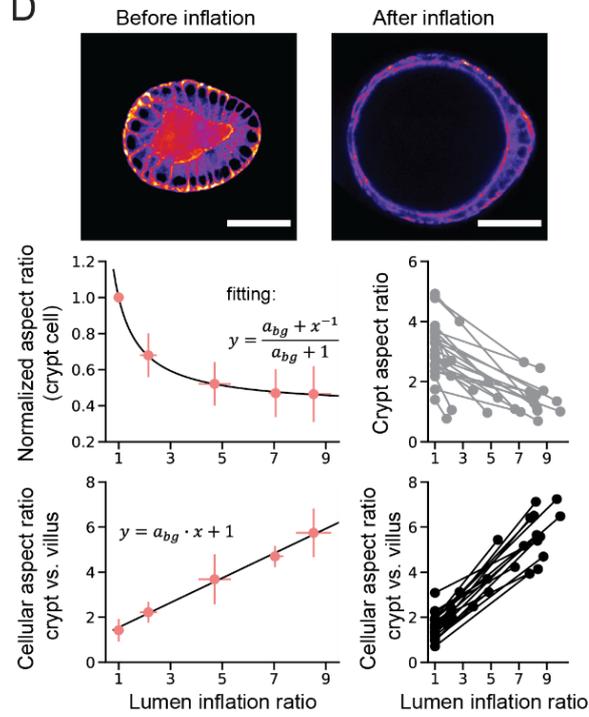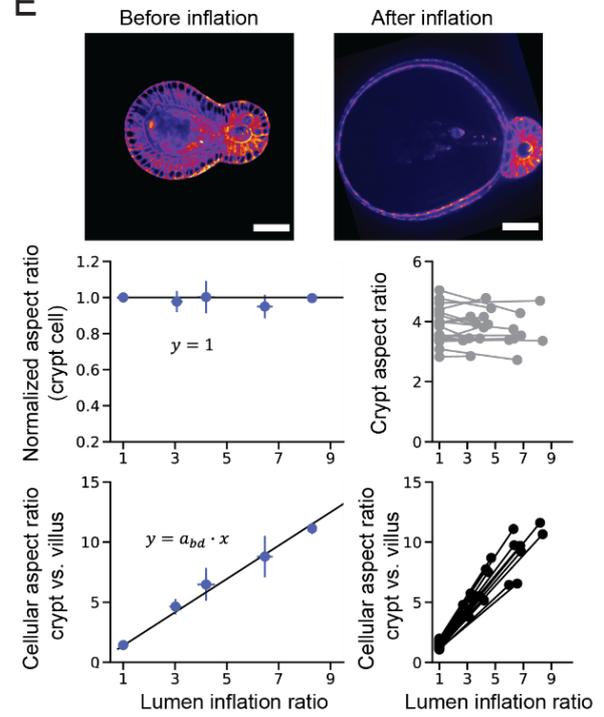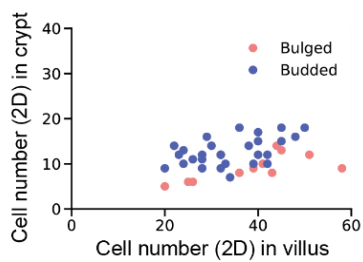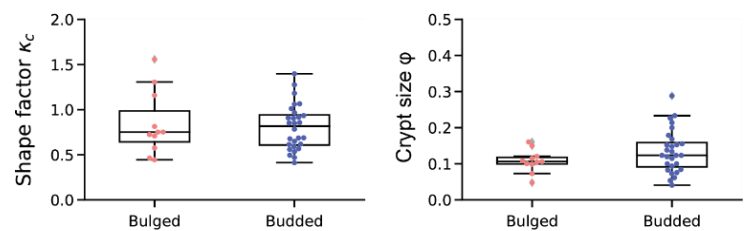

**Figure S7: Quantification of mechanical parameters. A**. Definitions of cellular aspect ratios $\left(\frac{h}{d}\right)_c$ and $\left(\frac{h}{d}\right)_v$, and in-plane contraction ratio $\alpha = \frac{(\Gamma_a+\Gamma_b)_c}{(\Gamma_a+\Gamma_b)_v}$, with subscripts c and v respectively denoting crypt and villus. **B**. Quantification of Myh-9-GFP intensity and cell shape: top, crypt to villus Myh-9-GFP intensity ratios, including crypt apical (ca) to villus apical (va), and crypt basal (cb) to villus basal (vb); bottom, quantification of cellular aspect ratios in crypt $\left(\frac{h}{d}\right)_c$ and that in villus $\left(\frac{h}{d}\right)_v$. **C**. Inference of in-plane contraction ratio $\alpha$ from both Myh-9-GFP intensity and cell shape: top, the ratio of in-plane Myh-9-GFP intensities (i.e. the sum of apical and basal intensities) of crypt to villus cell; bottom, the aspect ratio of crypt to villus cell $\left(\frac{h}{d}\right)_c / \left(\frac{h}{d}\right)_v$. In panel B and C, both bulged ($N = 11$) and budded ($N = 28$) samples are quantified. Bulged samples with lumen breakage (labelled as $v \approx 1$, also used in Fig. 2E) are chosen to quantify cellular aspect ratio as their cell shapes give direct inference of $\alpha$ (see SI Note for details). **D, E**. Fitting of the lumen inflation-induced morphometric variation with analytic formulae derived from the model. The images (top) show the organoid morphology and cell shape before and after lumen inflation by PGE treatment. Morphometric variation in cells is tracked by (middle) normalized aspect ratio of crypt cell $\left(\frac{h}{d}\right)_c^{ifl} / \left(\frac{h}{d}\right)_c^{ctr}$ (i.e. the value after lumen inflation $\left(\frac{h}{d}\right)_c^{ifl}$ normalized by that before inflation $\left(\frac{h}{d}\right)_c^{ctr}$, data of $\left(\frac{h}{d}\right)_c^{ctr}$ and $\left(\frac{h}{d}\right)_c^{ifl}$ shown in the right panel), and (bottom) aspect ratio of crypt to villus cell $\left(\frac{h}{d}\right)_c / \left(\frac{h}{d}\right)_v$ (original data shown in the right panel), in both bulged (panel D, $N = 24$) and budded (panel E, $N = 19$) organoids. **F.** Quantification of cell numbers in crypt and villus cross-sections (2D) from organoid images, which are used to estimate parameters in panel G (see SI Note for details). **G**. Measurement of shape factor $\kappa_c$ (left) and crypt size $\varphi$ (right) in bulged ($N = 11$) and budded ($N = 28$) crypts. Scale bars: 50 µm. Error bars represent SD.

**Supplementary Video 1. Representative light-sheet time-lapse recording of normal crypt formation.**

A full stack of an organoid expressing LifeAct-GFP is acquired every 20 minutes for 16:30 hours from Day3 till budding. Left panel, single plane intersecting the middle of the organoid. Right panel, maximum z-projection. Experiments were repeated with at least five independent recordings. Scale bar, 50 µm.

**Supplementary Video 2. Representative light-sheet time-lapse recording of lumen inflation in bulged organoid**

A full stack of budded organoid expressing LifeAct-GFP is acquired every 30 minutes from Day3.5 bulged organoid, Right panel, maximum z-projection. Experiments were repeated with at least five independent recordings. Scale bar, 50 μm.

**Supplementary Video 3. Representative light-sheet time-lapse recording of lumen inflation in budded organoid**

A full stack of an organoid expressing LifeAct-GFP is acquired every 3 minutes for around 2 hours from Day4 budded organoid. Left panel, single plane intersecting the middle of the organoid. Right panel, maximum z-projection. Experiments were repeated with at least five independent recordings. Scale bar, 50 μm.

**Supplementary Video 4. Representative spinning-disc time-lapse recording of Myh9-GFP relocation in bulged organoid treated with lumen inflation.**

A full stack of bulged organoid expressing Myh9-GFP is acquired when the change of the organoid lumen volume was notable by eye. The organoid was treated with 0.5 μM PGE, cultured and recorded for 1.5 hours. Experiments were repeated at least three times. Scale bar, 50 μm.